# A magnetic liquid deformable mirror for high stroke and low order axially symmetrical aberrations.

**Denis Brousseau, Ermanno F. Borra, Hubert-Jean Ruel and Jocelyn Parent**

*Département de physique, génie physique et optique and Centre d'Optique Photonique et Laser (COPL), Université Laval, Québec, Québec, Canada G1K 7P4*
*denis.brousseau.1@ulaval.ca, borra@phy.ulaval.ca, hubert.jean-ruel.1@ulaval.ca, jocelyn.parent.1@ulaval.ca*

**Anna Ritcey**

*Département de chimie and COPL, Université Laval, Québec, Québec, Canada, G1K 7P4*
*anna.ritcey@chm.ulaval.ca*

**Abstract:** We present a new class of magnetically shaped deformable liquid mirrors made of a magnetic liquid (ferrofluid). Deformable liquid mirrors offer advantages with respect to deformable solid mirrors: large deformations, low costs and the possibility of very large mirrors with added aberration control. They have some disadvantages (e.g. slower response time). We made and tested a deformable mirror, producing axially symmetrical wavefront aberrations by applying electric currents to 5 concentric coils made of copper wire wound on aluminum cylinders. Each of these coils generates a magnetic field which combines to deform the surface of a ferrofluid to the desired shape. We have carried out laboratory tests on a 5 cm diameter prototype mirror and demonstrated defocus as well as Seidel and Zernike spherical aberrations having amplitudes up to 20 μm, which was the limiting measurable amplitude of our equipment.

## 1. Introduction

Adaptive optics are increasingly used for optical applications [1]. Conventional deformable mirrors use solid thin plates or membranes [2,3,4]. They are expensive, have few actuators, are typically limited to diameters of a few cm and strokes of a few microns, a major limitation for some applications such as vision science. Liquid mirrors offer an interesting alternative that does not suffer from these restrictions. For example, they can have smooth deviations from flatness measuring several centimeters.

The fabrication of liquid optical components is recent but such elements are presently used and are the subject of intensive research and development work. The fundamental advantage of liquid optics comes from the fact that the surface of a liquid follows an equipotential surface to a very high precision. This has been used to make large parabolic mirrors by rotating containers filled with mercury. Optical shop tests [5,6], have shown that rotating mercury mirrors have excellent optical qualities. Liquid Mirror Telescopes (LMTs) have been built and used to obtain astronomical, space science and atmospheric sciences data [7-11]. The main advantage of those liquid mirrors resides in their low costs (about two orders of magnitude less than conventional parabolic glass mirrors). Liquid adaptive mirrors shaped by electrocapillary effects have also been discussed [12].

In this article, we discuss liquid surfaces shaped by magnetic fields. Mercury is poorly-suited to make liquid magnetic mirrors since ferromagnetic mercury is unstable [13], and the high density of mercury necessitates strong magnetic fields and thus high electric currents [14,15]. These difficulties can be solved by using a ferrofluid coated with a metallic layer [16]. Ferrofluidic mirrors that use arrays of actuators have been previously discussed [17]. This technology has been substantially improved in our laboratory since that article and new results will be presented elsewhere. Although these ferrofluidic mirrors can have substantially larger strokes than solid or membrane mirrors, they cannot, in practice, give extremely large useful strokes because the footprint of the individual actuators would lead to surfaces having unacceptable granularity. This problem can be greatly reduced by generating a magnetic field with simple networks of wires. In this article we present numerical simulations and experimental results of ferrofluidic mirrors shaped by a simple geometrical arrangement of concentric rings of current carrying wires.

## 2. Ferrofluid mirror technology

Ferrofluids are liquids that contain a suspension of colloidal ferromagnetic particles within a carrier liquid. In the presence of an external magnetic field, the magnetic particles align themselves with the field and the bulk of the liquid becomes magnetized. Energy conservation considerations lead to a surface governed by magnetic, gravitational and surface tension forces. Stable ferrofluids having wide ranges of physical properties can be produced to suit each practical application needs. The amplitude of a deformation, generated by an applied magnetic field, on a ferrofluid surface, can be approximated by [18]

$$h = \frac{\mu_0(\mu_r - 1)}{2\rho g}\left(\mu_r H_n^2 + H_t^2\right) \qquad (1)$$

Where $\rho$ is the density of the ferrofluid and $H_{n,t}$ are the normal and tangential components of the magnetic field intensity within the ferrofluid. The relative permeability $\mu_r$ is assumed to be independent of the magnetic field since in most cases it is extremely small compared to the saturation magnetization of the ferrofluid. From this and using the usual magnetic boundary conditions, we derive the formula for the amplitude of a single peak in terms of the external applied field

$$h = \frac{(\mu_r - 1)}{2\mu_r\mu_0\rho g}\left(B_n^2 + \mu_r B_t^2\right) \qquad (2)$$

The amplitudes of the surface deformations are limited by the Rosensweig instability which occurs when the magnetic field exceeds a critical value, generating a series of spikes, visible to the naked eye [19]. This critical value depends on the physical parameters of the ferrofluid and the normal component of the magnetic field relative to the surface. For typical ferrofluids and the worse-case geometry (the field is entirely perpendicular to the surface), the Rosensweig instability occurs around 80 Gauss. Using Eq. (2) to obtain the maximum surface deformation before the onset of the instability, we get an amplitude of over a millimeter. However, in theory, much larger deformations (several tenths of cm) can be obtained for fields that have components mostly parallel to the surface [20].

Ferrofluids have low reflectivity and for many applications must be coated with a reflective layer. This can be done with nanoengineered reflective liquids based on interfacial films of silver particles known as Metal Liquid-Like Films or MELLFs [21,22,23]. MELLFs combine the properties of metals and liquids, can be deformed and are therefore well adapted to applications in the field of liquid optics. The fabrication of a nanoengineered liquid mirror has been described elsewhere [24] .This technology is young and can certainly be improved. We are presently carrying out work to improve the properties of MELLF-coated liquids. We have also successfully coated other liquids with conventional vacuum-coating equipment.

## 3. Numerical simulations and experimental results

*3.1 Numerical simulations*

The cylindrical symmetry of most optical components suggests that the magnetic field should be produced with electric currents flowing in circular coils. We thus computed numerically the magnetic field generated by a concentric array of circular coils using the Biot-Savart law and used Eq. (2) to estimate the parametric surface of the ferrofluid.

An equally spaced array of 3 concentric coils was our first guess to produce axially symmetrical aberrations on the ferrofluid surface. To find optimal currents, we carried out Monte Carlo simulations that use an adapted version of the well-known downhill simplex method in multi-dimensions described in Numerical Recipes in C [25]. The downhill simplex method is due to Nelder and Mead [26], and requires only function evaluations, not derivatives. It is not very efficient in terms of the number of function evaluations it requires, but the computational burden is small and it permits to get results fast. To minimize computer time, we assumed that each coil is made of circular loops of vanishingly thin wire. The function we are optimizing is the standard deviation between the actual calculated surface and the desired one. The algorithm stops when the deviation reaches a user-set value. The starting points are randomly generated within a null and a maximum desired current. It was soon found that 3 coils were enough to give a defocus term with low residual but not for a primary spherical aberration. We thus increased the number of coils to 5 and were then able to get low residuals on both terms. Height of the coils relative to each other was also optimized. A compromise between optimized vertical positions and design simplicity was finally chosen.

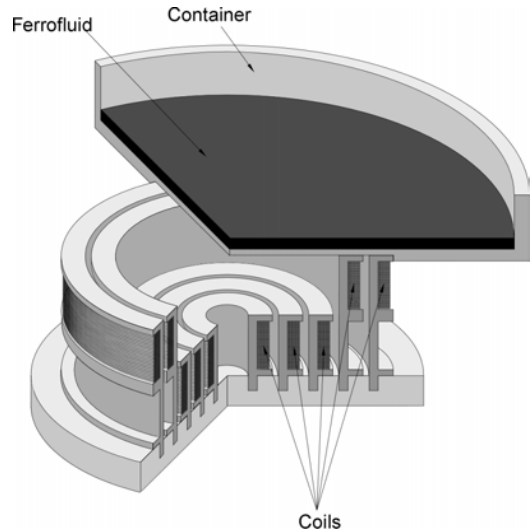

Fig. 1. Schematic of the mirror. The outer coil has a diameter of 10 cm.

*3.2 Experiments*

We carried out experiments to test the validity of our simulations. They were done with a mirror that uses 5 circular concentric coils with diameters ranging from 2 to 10 cm made of copper-wire windings. For each coil, we used 162 windings, except for the central one which has 108, of AWG 26 gauge copper wires laid over machined aluminum cylinders. The coils are easy to build and lend themselves to low cost mass production. Fig. 1 shows a schematic drawing of the actual mirror. The mirror is surrounded by a cylindrical magnetic shielding enclosure (not shown in Fig. 1) to isolate it from surrounding magnetic fields (mainly from the Earth). A circular hole allows the beam to reach the surface of the mirror. To avoid the meniscus contribution to the final surface along the edges, the container holding the liquid must be larger than the surface to be sampled by a few centimeters. The ferrofluid used was EFH1 from Ferrotec Corp. and has a density of 1.21 g/cm$^3$, a magnetic permittivity of 2.7 and a viscosity of 6 cp.

To test the surfaces, we used a general purpose Mach-Zehnder interferometer (GPI) from Zygo, as well as a Shack-Hartmann (SH) wavefront sensor from Imagine Optics. Ferrofluidic surfaces can have very large deformations, but because of limits set by the interference fringe spacing, we could only measure deformations of a few μm with the GPI. The Shack-Hartmann sensor can measure peak-to-valley deformations as large as 20 μm.

We found that the agreement between the theoretical aberration terms and the surfaces obtained experimentally by applying the computed currents to the coils was poor, with errors of the order of a few tens of percents. We suspect that the problem arises from the approximations used to derive Eq. (1) and Eq. (2). Interaction between a ferrofluid and a magnetic field is very complex as volume and surface effects play a major role in the final equilibrium state of the liquid. One cannot determine the exact surface of a ferrofluid without FEM calculations, which is not desirable in a closed loop AO system. We thus had to manually tweak the currents to improve the agreement. With the corrected currents we obtained an excellent agreement with the corresponding aberrations as demonstrated in Fig. 2 that compares theoretical and experimental wavefronts for defocus as well as Zernike and Seidel spherical aberrations. The opposite sign counterpart of each of these terms have also been obtained. The histograms on the right show that the desired aberration term does indeed dominate. It must also be noted that the adjusted currents are probably not optimal and that a feedback loop algorithm should further improve the agreement. We found that the system is stable in time. The experimental wavefronts can be recreated by applying the same currents over several weeks. We also found that the amplitudes of the wavefronts were scalable by applying a power scale law as seen in Fig. 3. Consequently, it appears that the numerical simulations give a useful first guess to the currents which can then be optimized. The optimized currents can then be used and amplitudes scaled with an empirical law as the one in Fig. 3.

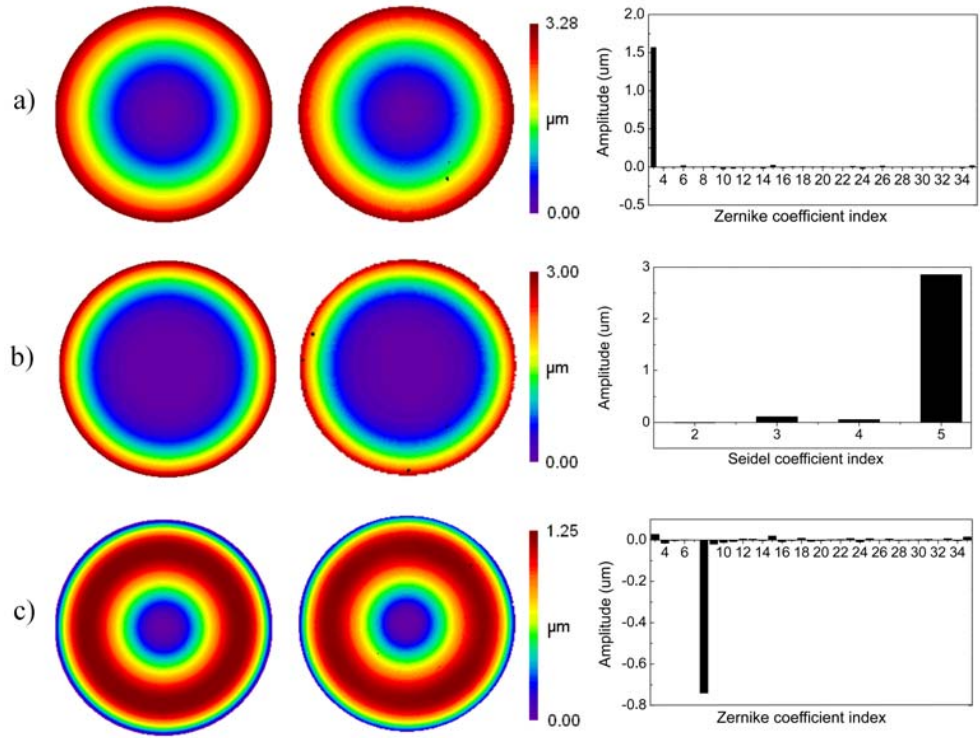

Fig. 2. Theoretical wavefronts (left), measured wavefronts (center) and aberration coefficients histogram (right). The Mach-Zehnder interferometer measurements shown are for the best experimentally achieved combination of currents. Illustrated terms are: a) defocus, b) Seidel spherical aberration and c) Zernike spherical aberration.

*3.3 Response time.*

The frame rate of the GPI is too slow to observe the dynamics of liquid surfaces. We were however able to obtain useful data with triggered imaging with the SH sensor. To have a rough estimate of the response time, we did a simple experiment by applying a 1 ms voltage pulse to a circular coil and then followed the surface response with the SH sensor. We found that the surface deviation varies from zero to a maximum height of 10 μm in less than 50 ms before falling back. This measure corresponds to the worst case scenario where the surface goes from rest to its maximum amplitude. In normal use of the mirror, only small variations around a bias surface are needed. Response time also depends on the physical properties of the ferrofluid such as viscosity, depth of the layer, and how the magnetic field is applied. We are presently studying response times in detail and investigating techniques that promise to improve them. For example, numerical simulations and simple experiments show that overdriving techniques should give better response times. We overdrive by applying to a coil a brief but very intense current. The results of ongoing laboratory investigations will be presented elsewhere when completed. It should also be possible to improve the response time by depositing a membrane on the surface of the mirror.

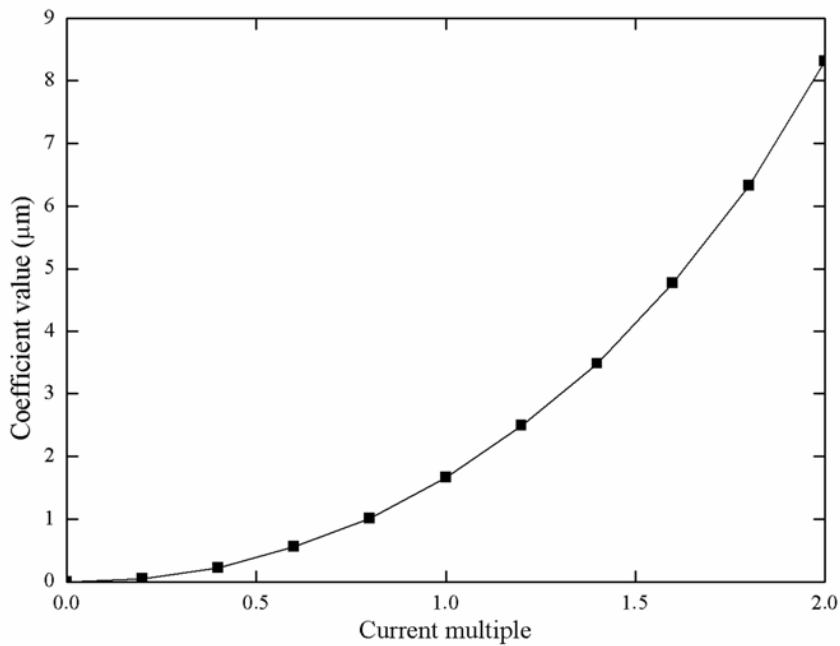

Fig. 3. Amplitude of the defocus term as a function of the current multiple. Current multiple of 1 corresponds to a) in Fig. 2. Even at the maximum amplitude of the defocus, remaining aberration terms are well below 5% and decrease as current multiple increases.

## 4. Discussion and conclusion

We have carried out numerical simulations that show that it is possible to generate axially symmetrical wavefront aberrations by applying electric currents to a few coils made of copper wire wound on aluminum cylinders. The coils generate a magnetic field that deforms the surface of a ferrofluid to the desired shape. We have carried out experiments on a small prototype mirror and demonstrated defocus as well as Seidel and Zernike spherical aberrations.

The currents required with the copper windings are reasonable (2 amperes maximum) if one only needs to generate deformations of the order of 20 microns, but become high if one wishes to shape optical elements, like adaptive primary or secondary mirrors, having peak-to-valley deviations from flatness of the order of a mm or greater. In that case, one would have to add ferromagnetic material or more windings, to intensify the field generated by the wires or use superconducting cables as noted in [20]. Superconducting materials capable of working at liquid nitrogen temperatures are commercially available. Because superconductor research is driven by practical applications, we can expect major advances over the next few years and the availability of superconducting wires at increasingly higher temperatures. A recent note in Scientific American, issue of August 2006, states that tens of kilometers of superconducting cables working at liquid nitrogen temperatures will soon be commercially available. Quantum wires, wires spun from carbon nanotubes, offer another promising technology for they conduct much better then copper and dissipate very little electricity as heat. They are the subject of considerable research efforts, driven by use in power grids. Unlike superconducting wires, there is no need to cool them. Thus, for applications using optical elements the size of our prototype and amplitudes of the order of 20 microns, power consumption and temperature problems are not significant. Active cooling may be necessary for amplitudes significantly larger than 20 microns. For extremely larges amplitudes (greater than a millimeter), it may be necessary to use superconducting wires. As stated above, this issue is discussed in [20].

Liquid magnetic optics presents advantages and disadvantages with respect to solid optics. The main advantages are large stroke, low cost and scalability. This technology could lead to extreme deformable optics capable of strokes of hundreds of µm and more. Very large deformable mirrors having diameters greater than a meter should be feasible. In a recent article [20], we discuss an extreme application of this technology to shape a 30 m diameter mirror having deviations from flatness of the order of a meter. Magnetic fields could also be used to make ferrofluidic concave or convex mirrors having arbitrary shapes. For example, by applying a sufficiently high current, one could make a 5 cm diameter parabolic mirror (or a spherical one) and then add an additional amount of spherical aberration as needed by a particular optical design. The radius of curvature of the mirror and the amount of aberration could vary in time. This could be useful for a wide range of applications like optical testing and vision science. It is well know that the human eye suffers from many optical imperfections. Ophthalmologists look into the eye to detect and diagnose diseases by examining the retina. The quality of the image they see is strongly dependant on the level of aberration in the eye. Previous studies in a large population [27], show that Zernike defocus accounts for 80% of the total variance of the wavefront aberration in the human eye and has also the largest amplitude. The amount of the higher-order aberrations generally declines with the order number, except for spherical aberration, which is larger in mean absolute rms value than any third order mode. Getting rid of these two axially symmetrical aberrations with liquid adaptive optics can therefore greatly improve the image we get from the retina in a very simple and affordable way. Also, imaging of the retina is often done at several different wavelengths, and defocus is directly proportional to the wavelength. The optical setup needs to be modified accordingly and this can be troublesome and time consuming. Being able to correct the defocus with a liquid deformable mirror would be of great benefit.

There also are disadvantages. The response time seems to be significantly slower than the response of solid mirrors. A simple experiment indicates response times of the order of 50 ms. This may be a problem for some applications (e.g. Astronomy) but not for others (e.g. Ophthalmology) [28]. As mentioned in section 3.3, we are currently investigating techniques that may significantly improve response time. The surface of a liquid is also sensitive to vibrations. Fortunately, the sensitivity to vibrations decreases significantly when a layer of liquid thinner than a millimeter is used [29]. A simple technique, illustrated by Fig. 13 in the paper from Borra et al. [29], allows having layer thicknesses less than the limit imposed by surface tensions. Furthermore, our experiments show that the sensitivity to vibrations is significantly reduced when a magnetic field is applied to the ferrofluid. Although care must be taken to protect magnetic liquid mirrors from vibrations, we find that they are not a major issue. We use a thickness of about two millimeter on a basic optical table with minimal damping, and find that the amplitude of disturbances induced by vibrations is less than 1/10 of a wave.

Prima facie, one would expect that a major limitation comes from the fact that the liquid surface must be horizontal and that it could not be tilted or used in space. However, this is not the case since the ferrofluids stick to magnets. As a quick experiment, we have put a MELLF-coated ferrofluid on a permanent magnet and turned it upside down. We did not quantitatively measure the surface but it looked smooth to the unaided eye. There may thus be an intriguing application of these mirrors in astronomical telescopes where a ferrofluidic adaptive secondary mirror could consist of a permanent magnet to which one adds electric wires and coils to shape the wavefront. Changing the inclination of such a mirror would induce a changing wedge into it. However, the wedge could be eliminated with an array of current-carrying wires or simply by mechanically tilting the primary mirror by an appropriate amount that cancels the wedge.

We cannot simply sum scalar deformations from individual coils to obtain the wavefront generated by the sum of each coil. This is because individual deformations are the result of summing vector fields and not scalar fields. The surface obtained with several coils is thus shaped by the sum of the magnetic vectors instead of the sum of surfaces generated from each winding. We have recently developed an efficient algorithm for the superposition of fields from a hexagonal structure of actuators of the type discussed in [30]. Hopefully, we will be able to adapt it to the superposition of fields from arrays of wires.

Ferrofluid deformable mirrors are unidirectional; the coils can push but not pull. Pulling is done by a combination of gravitational forces and hydrodynamic flow generated among working coils. To achieve the push-pull effect required for normal wavefront control, the mirror must be first biased by driving the coils to produce a piston term of about 50% of their maximum deflection; pulling is done by lowering the current in a given coil. Also, considering that the surfaces are formed by vectorial addition of the magnetic field components, one can use this to advantage by reversing the currents in some of the coils to give a wider range of possible surfaces.

Finally, let us mention that we are presently working on complex liquid surfaces shaped by conducting wire networks of arbitrary shapes. They can be used to generate wavefronts having non-cylindrical symmetries (e.g. coma and astigmatism). This opens the possibility of generating higher order wavefronts by adding the magnetic fields from differently shaped wire networks.

Ferrofluidic mirrors shaped with wire networks are unlikely to be able to generate the complex shapes feasible with mirrors that use individual actuators. However, their stroke advantage can make them useful for applications where actuator-driven mirrors cannot compete. For example, a 5 cm diameter F/1 mirror having a spherical surface with added aberrations could be shaped and used in an optical system to correct the aberrations of the system. This could yield a much simpler system than the complicated succession of mirrors and lenses that must often be used. The fact that this kind of mirror can dynamically change shape could also be used to advantage. Another example of application is their use as reference wavefronts. In optical shop tests, one often has to subtract a spherical reference wavefront. For that purpose, the Zygo GPI uses costly auxiliary lenses. Reference wavefronts of different curvatures need different lenses having discrete radii. A single liquid deformable mirror would yield an infinite number of surfaces of continuously variable curvatures. For many applications this mirror would not need a reflective coating (e.g. testing mirrors and lenses).

The bottom line is that ferrofluidic mirrors are a new type of versatile and inexpensive optical elements that offer significant advantages over solid optics. Although the initial design is very crude, it is able to give low order aberration terms with large amplitudes and good surface quality at a very low cost. This very simplicity is one of the major advantages of this technology.

**Acknowledgments**

This research was supported by the Natural Sciences and Engineering Research Council of Canada.